\documentclass[12pt]{article}

\title{\bf Simplicissimus' Inquiries}
\author{Vadim A. Kuzmin \\ {\it Institute for Nuclear Research of the Russian
Academy of Sciences,} \\ {\it 60th October Anniversary Prospect
7a, Moscow 117312, Russia}}
\date{}
\begin{document}
\maketitle

\hspace*{6cm}   {\bf  Dedicated to:}\\
\hspace*{6cm}   {\bf Hans Jakob Grimmelshausen}\footnote{H. J. Grimmelshausen (1621-1676), the author of the novel ``Simplicissimus'' (1668), M.,Khudozhestvennaya literatura. 1976.},\\
\hspace*{6cm} {\bf Nikolay Nikolaevich Bogolyubov},\\
\hspace*{6cm}  {\bf Yakov Borisovich Zeldovich}     

\bigskip

This is a kind of essay dedicated to Lev Borisovich Okun.

I would like to ask some questions which I can't answer. However I
suspect I know how to answer them. Questions are as follows:
\begin{enumerate}
\item Particle identity. \item Information storage. \item
Information read-out at the speed higher than that of light.
\end{enumerate}

\section{Particle Identity}

Somehow all interactions of particles obey the same laws in
Universe both at Alfa-Centauri and everywhere, and even in INR.

This is surprising and even mysterious.

Electrons are identical everywhere.

What is the reason of this phenomena?

Any replication might have mistakes with some probability. We
guess it should be about $10^{-40}$ . This might arise because of
a small violation of the Pauli principle.

In genetics we know the examples of identities - children are
similar to their parents, because they are copies of the parents
on the cell level. Why the electrons are identical  to each other?
Because they are copies of their parents.

\section{Information Storage}

All interactions are always and everywhere identical. Why is it
so? If we want to store any information it should be saved
somewhere in discrete structures, because information can't be
saved on a continuous set. We have to mention here the cellular
automata.

Information should be stored somewhere.

Where? In electron or in medium.

Certainly, it is obvious that an electron is unable to store the
information. Then the only choice is medium. As a store room is
medium we estimate the number of cells as $10^{80}$.

In case electrons do contain this information, I will address the
same questions to their components. And so on.

{\bf Vacuum is the only medium capable to store the necessary
information.}

\section{Speed Higher than that of Light}
                                                                                                                                 
A photon has regular behavior. Photons time of flight through the
target is about $10^{-13}$ sec. It is hard to imagine that a
particle might get any information during this period of time --
what it is and how should it behave.

{\bf So the speed of communication should be certainly greater than
the speed of motion. Consequently particles should read out the
information at the speed higher than that of light.}

\section*{Acknowledgements}

I am very much obliged to S.A. Demidov, A.U. Ignatyev, V.A. Matveev, V.A.
Rubakov, A.N. Tavkchelidze, I.I. Tkachev for useful discussions
and support.

\end{document}